\documentclass[aps,prd,showpacs,nofootinbib]{revtex4}

\usepackage{amsfonts}
\usepackage{verbatim}
\usepackage{amsmath}
\usepackage[USenglish]{babel}
\usepackage[dvips]{epsfig}
\usepackage{graphicx}
\usepackage{float}
\usepackage{epsfig,subfig}
\usepackage{epstopdf}
\usepackage{hyperref}
\usepackage{cleveref}
\usepackage{color}
\DeclareMathAlphabet\mathbfcal{OMS}{cmsy}{b}{n}
\usepackage{lipsum}

\begin{document}

\title{On the equivalence between bumblebee models and electrodynamics in a non-linear gauge}

\author{C. A. Escobar}
\email{cruiz@ualg.pt}
\affiliation{CENTRA, Departamento de F\'isica, Universidade do Algarve, 8005-139 Faro, Portugal.}

\author{A. Mart\'{i}n-Ruiz}
\email{alberto.martin@nucleares.unam.mx}
\affiliation{Instituto de Ciencias Nucleares, Universidad Nacional Aut{\'o}noma de M{\'e}xico, 04510 M{\'e}xico, Distrito Federal, M{\'e}xico.}

\begin{abstract}
Bumblebee models are effective field theories describing a vector field with a nonzero vacuum expectation value that spontaneously breaks Lorentz invariance. They provide an alternative way of exploring the similarities between theories with spontaneous Lorentz symmetry breaking and gauge theories. The equivalence between bumblebee models with suitable conditions and standard electrodynamics in a non-linear gauge $A_\mu A^\mu-b^2=0$ is taken for granted; however, this point is very subtle and has not yet been fully addressed. The main goal of this paper is to fill in this gap. More precisely, here we study the relation between a bumblebee model, with a smooth potential of the form $V(B _{\mu}) = V (B _{\mu} B ^{\mu} - b ^{2})$, and standard electrodynamics in the non-linear gauge $A _{\mu} A ^{\mu} - b ^{2} = 0$, both at the classical and quantum levels. Using the Dirac's method we show that after introducing Dirac brackets with suitable initial conditions, the classical dynamics of the bumblebee model corresponds to that of standard electrodynamics in the aforementioned non-linear gauge. In the quantum case we demonstrate that perturbative calculations of Feynman amplitudes to any physical process in each model are indistinguishable. To do this, we show that the Feynman rules and propagators of standard electrodynamics in the non-linear gauge and those describing the bumblebee model are the same.
\end{abstract}

\pacs{11.15.-q, 11.30.Cp, 12.20.-m}
\maketitle

\section{Introduction}

Recently, Lorentz violation (LV) has attracted great attention from both theoretical and experimental sides. Strong motivation exists to study this subject. On the theoretical front, some modern approaches suggest that Lorentz invariance can be broken at high energies; and from the experimental point of view, since Lorentz symmetry plays a fundamental role in our current and successful theories (such as the Standard Model of particles physics and General Relativity), it must be experimentally tested. Examples of the former include quantum gravity \cite{Others}, string field theory \cite{string}, brane-world scenarios \cite{brane}, noncommutative field theories \cite{noncumative}, condensed matter analogues of ``emergent gravity'' \cite{emergentG}, doubly special relativity \cite{DSR}, aether theories \cite{aether}, four-dimensional spacetimes with a nontrivial topology \cite{topology}, varying speed of light cosmologies \cite{cosmology} and emergent gauge bosons \cite{gaugeb}, just to name a few.

Among the schemes considering spontaneous Lorentz symmetry breaking we can mention the so-called bumblebee models \cite{Bumble1,Bumble2}, which are vector field theories. They have been extensively studied in curved and flat spacetimes \cite{Bumblstudy1,Bumblstudy2,Bumblstudy3,radiativecorrect,HERNASKI,Potting1}, including astrophysical and cosmological contexts \cite{bumbleAstro,bumbleCosmo}. Bumblebee models are not gauge invariant and are defined from different forms of the kinetic and the potential terms for the vector field, also the coupling with matter and gravity can be constructed in different ways \cite{Potting1}. A relevant feature of bumblebee models is that these allow a detailed description of the spontaneous Lorentz symmetry breaking, for example, the vacuum condition together with the Goldstone bosons (GB) can be explicitly determined. Likewise, it has been studied the possibility of looking these Goldstone bosons in bumblebee models as an alternative to photons in a gauge theory \cite{Potting1}.   

Usually, the origin of massless particles is associated with gauge symmetry; in particular, the masslessness of the photon and graviton is explained by the U$(1)$ gauge invariance and diffeomorphism invariance, respectively. However, spontaneous symmetry breaking (SSB) can provide an alternative way to understand massless particles as Goldstone bosons. This proposal dates back to the earliest works of Bjorken and Guralnik \cite{Bjorken,Guralnik} where the pion interactions in the non-linear sigma model was characterized by spontaneous chiral symmetry breaking. After these works, Nambu proposed to describe photons as GB arising from a SSB of Lorentz invariance \cite{nambu}. Bumblebee models also provide an alternative way of exploring the similarities between theories with spontaneous Lorentz symmetry breaking and gauge theories. The main challenge posed by this setting is to show the conditions under which the violations of Lorentz symmetry, arising from a SSB, are unobservable, such that the Goldstone bosons that appear can be interpreted as the gauge particles of a standard gauge theory. In other words, one tries to determine the conditions under which the corresponding Lorentz-violating model is equivalent to a standard gauge theory.

In this regard, the conditions for the equivalence between Nambu models and standard electrodynamics, Yang-Mills theory or gravitation have been studied in Refs. \cite{parametri12,nambu1,Nambuphotons}. Conversely, the equivalence between bumblebee models and gauge theories has not been fully addressed. In Ref. \cite{Potting1} the authors performed a classical Hamiltonian analysis of bumblebee models to study their stability, equations of motion, degrees of freedom and constraints. They also worked out on the relation of these models with standard electrodynamics (ED) in a non-linear gauge when the phase space is restricted to suitable initial conditions for the former. However, even at classical level, some issues should be addressed in order to prove the equivalence. Particularly, in Ref. \cite{Potting1} was not studied the standard electrodynamics in the proposed non-linear gauge. One of the main points requiring a further discussion is related to the character of the conditions in both models, on the one hand we will find a gauge fixing condition in standard electrodynamics, while on the other hand, in the bumblebee model we will find a suitable initial condition which preserves, via the bumblebee dynamics, its value for all the time. Since these conditions are of different nature, even when they have the same form, they are treated in different way; therefore, it is not straightforward to establish that the contributions arising from these conditions are the same in both models. Mainly, the consistency in both theories has to be verified. The equivalence at the quantum level, which was not explored in Ref. \cite{Potting1}, requires a similar analysis on the conditions in both models.

Unlike electrodynamics, the introduction of a potential term in the bumblebee models drastically modifies the structure and dynamics of the theory, so that the equivalence between these theories can not be established in a simple way. To mention a few significant differences between a particular bumblebee model with the smooth quadratic potential $V (B _{\mu}) = \frac{\kappa}{4}(B _{\mu} B ^{\mu} - b ^{2} ) ^{2}$ and standard electrodynamics are: (i) the bumblebee model only has second-class constraints (there is no gauge invariance), (ii) the number of degrees of freedom (d.o.f.) in the bumblebee model is three, while ED has only two d.o.f., (iii) the equations of motion do not match and (iv) the current density conservation does not hold in the bumblebee model.

The main goal of this paper is to study the relation between a bumblebee model, with a smooth potential of the form $V(B_\mu)=V(B_\mu B^\mu-b^2)$, and standard electrodynamics in the non-linear gauge  $A _{\mu} A ^{\mu} - b ^{2} = 0$, both at the classical and quantum levels. The novel feature of this work, with respect to that in Ref. \cite{Potting1}, is that we complete the classical Hamiltonian analysis for the bumblebee model and perform the study of standard electrodynamics in the aforementioned non-linear gauge. We also extend the results for more general bumblebee potentials and study the equivalence from the quantum approach.

 We follow the usual Dirac's method \cite{Dirac} to analyze both theories classically. Imposing suitable initial conditions to the bumblebee model, we construct the Dirac brackets to analyze its consistency with standard electrodynamics in a non-linear gauge. The quantum equivalence is more subtle, and we demonstrate their equivalence from a perturbative perspective. Once the required conditions for the bumblebee model are introduced, we show that the propagators and Feynman amplitudes for physical processes arising from the bumblebee model and those stemming from the gauge-fixed standard electrodynamics are the same. Therefore, perturbative calculations of any physical process in each model can be seen to be indistinguishable, proving in this way their equivalence.

The outline of this work is as follows. In Sec. \ref{BumblebeeSection} the bumblebee model is introduced and some of its main properties are reviewed. We also discuss the conditions which guarantee the equivalence between these bumblebee models and standard electrodynamics in the non-linear gauge. In Sec. \ref{parteclasica} we first work out their relation at classic level, while their equivalence at quantum level is studied in Sec. \ref{partecuantica}. Our summary and conclusions are contained in Sec. \ref{conclusiones}.

Here, Lorentz-Heaviside units are assumed ($\hbar =c=1$), the metric signature will be taken as $\left( +,-,-,-\right) $, and we use the conventions, Greek indices $\mu,\nu=0,1,2,3$, and Latin indices $i,j=1,2,3$.


\section{The bumblebee model}
\label{BumblebeeSection}

The bumblebee Lagrange density is defined by
\begin{equation}
\mathcal{L}_B(B_\mu)=-\frac{1}{4} B_{\mu\nu} B^{\mu\nu}-\frac{\kappa}{2}V(\xi)-B_\mu J^\mu,
\label{LagrangianBumblebee}
\end{equation}
where $\xi \equiv B_\mu B^\mu- b^2$, with $b^2$ a positive constant with dimensions of $[$mass$]^2$,  $B_{\mu\nu}=\partial_\mu B_\nu-\partial_\nu B_\mu$, $\kappa$ a dimensionless positive constant  and $J^\mu$ an external current. Gauge invariance is lost due to the potential term. 

The equations of motion arising from the Lagrangian density (\ref{LagrangianBumblebee}) are 
\begin{equation}
\partial_\mu B^{\mu\nu}-\kappa V'(\xi)B^\nu=J^\nu ,
\end{equation}
where the prime denotes derivate with respect to $\xi$. From the above equation we can see that in the bumblebee model the conservation of the current $J^\nu$ does not follow as a consistency condition from the equations of motion, as happens in standard electrodynamics.

Hereafter we only consider potentials satisfying the condition $V(\xi=0)=0$. We also assume that the potential  $V(\xi)$ has a degenerate minimum with respect to its argument, i.e. $V ^{\prime} (\xi=0)=0$. This implies that the potential is constrained to be zero by the relation $(B_\mu B^\mu -b^2)|_{\mbox{min}}= 0$ in its minimum. This condition is satisfied when the bumblebee field has a nonzero vacuum expectation value (VEV)
\begin{equation}
\langle B_\mu\rangle=\mathcal{B}_\mu,
\label{vacio1}
\end{equation} 
where $\mathcal{B}_\mu \mathcal{B}^\mu=b^2$. Usually the bumblebee potentials are taken to be of the form
\begin{equation}
V(B_\mu)=\frac{(B^\mu B_\mu-b^2)^n}{n}, \quad\quad\quad n\geq 2,
\label{potential12}
\end{equation} 
where $n=2$ is the case most frequently used. Clearly the potential (\ref{potential12}) satisfies the required conditions $V (\xi=0) = V ^{\prime} (\xi=0) = 0$. It can also be considered more general potentials provided the  aforementioned conditions are satisfied, for example, any potential which can be written as a polynomial in powers of Eq. (\ref{potential12}).

Notice that the vector $\mathcal{B}_\mu$ in Eq. (\ref{vacio1}) defines a preferred direction in spacetime, yielding to the spontaneous symmetry breaking of Lorentz invariance. The Goldstone theorem \cite{GoldstoneT} predicts the appearance of massless particles. However, due to the bumblebee potential, which implies the lack of gauge invariance, massive particles can also arise in these models \cite{Potting1}. We can find the Goldstone bosons from those Lorentz generators $G_\mu\,^\nu$ that do not leave the vacuum invariant $\delta_G \mathcal{B}_\mu=G_\mu\,^\nu \mathcal{B}_\nu\neq0$. Massive modes should oscillate perpendicularly to these GB.

Given a basis of Lorentz transformation generators $\mathcal{G}=\{G_i\}$ with $i=1,2,...,6$, it may happen that the number of $G_i$ satisfying the relation $\delta_{G_i} \mathcal{B}_\mu\neq0$ is different depending of the chosen vacuum. For a nonzero $\mathcal{B}_\mu$, from the calculation of $\{\tilde{c}_i\}=\{\delta_{G_i} \mathcal{B}_\mu$ \& $i=1,...,6$\} we can expect to find one to six nonzero four-vectors; however, it can be proved that the dimension of the spanned space by the set $\{\tilde{c}_i\}$ is three. This agrees with the particular cases in Ref. \cite{Carroll} when the spontaneous Lorentz symmetry breaking is performed from the six-dimensional space $SO(1,3)$ to the three-dimensional subspaces $SO(3)$ and $SO(1,2)$, corresponding to $\mathcal{B}_\mu$ timelike and spacelike, respectively, and where the standard generators of Lorentz symmetry have been considered to calculate $\{\tilde{c}_i\}$.

The above reasoning implies that the number of GB contained in the bumblebee model is three. Notice that, considering Lorentz invariance as a symmetry of the Lagrange density, this result is valid to any model in which the basic variable is a four-vector field with a nonzero vacuum expectation value, which is the case for most of the bumblebee models in flat spacetime.

It is worth to mention that there is not a continuos limit of this model to standard electrodynamics taking $\kappa\rightarrow0$. For any $\kappa\neq0$ the canonical structure and dynamics of the bumblebee model differs from that of ED. 

Next we review the main properties of the bumblebee model and point out the similarities and differences with standard electrodynamics. Since the potential in the Lagrange density Eq. (\ref{LagrangianBumblebee}) does not involve velocities, the standard canonical momenta are unaffected  
\begin{equation}
\frac{\partial \mathcal{L}_B}{\partial\dot{B}^0}=\Pi^0=0,\quad\quad\quad\frac{\partial \mathcal{L}_B}{\partial\dot{B}_j}=\Pi^j=\partial_0 B_j- \partial_j B_0.
\end{equation}
The canonical bumblebee Hamiltonian density $\mathcal{H}_c^B=\dot{B}_i\Pi^i-\mathcal{L}_B$ is
\begin{equation}
\mathcal{H}_c^B=\frac{1}{2}(\Pi^j)^2+\frac{1}{4}(B_{jk})^2-B_0(\partial_i \Pi^i-J^0)+\frac{\kappa}{2}V(\xi)+B_i J^i,
\label{hamiltoniano1}
\end{equation} 
with the canonical algebra given by the Poisson brackets 
\begin{equation}
\{B_\mu, B^\nu\}=0,\quad\quad,\{\Pi_\mu, \Pi^\nu\}=0,\quad\quad \{B_\mu, \Pi^\nu\}=\delta_\mu \,^\nu.
\end{equation}
Following Dirac's method for systems with constraints, two second-class constraints are identified 
\begin{equation}
\phi_1=\Pi^0,\quad\quad\quad\phi_2=\partial_j\Pi^j-\kappa B_0V'(\xi)-J^0.
\label{constricciones}
\end{equation}
The above implies that the number of d.o.f. of the bumblebee model is three  ($\#$d.o.f. $=\frac{1}{2}[$variables in the phase space - second-class constraints - 2$\times$first-class constraints$]=\frac{1}{2}[8-2-2\times0]=3$). This number is larger than that of the corresponding standard electrodynamics, which can be understood as a consequence of the lost gauge invariance. The equivalence between both theories requires at least to specify an additional condition in the bumblebee model to cut this extra degree of freedom.

To analyze the stability of the model, we rewrite the canonical Hamiltonian density (\ref{hamiltoniano1}) with $J^i=0$ in the form
\begin{equation}
\mathcal{H}_c^B=\frac{1}{2}(\Pi^j)^2+\frac{1}{4}(B_{jk})^2-\kappa B_0^2V'(\xi)+\frac{\kappa}{2}V(\xi),
\label{hamiltoniano1establ}
\end{equation} 
which is not positive definite over full phase space due to the terms involving the potential.

The time evolution of the canonical variables $B_\mu$ and $\Pi^\nu$, with $J^i\neq0$, is given by 

\begin{equation}
\dot{B}_j=\Pi^j+\partial_j B_0,\quad\quad\quad \dot{\Pi}^j=\partial_k \partial_k B_j-\partial_j \partial_k B_k+\kappa B_j V'(\xi)-J^j,
\label{equmov1}
\end{equation}
\begin{eqnarray}
\nonumber
&&\dot{B}_0=\frac{1}{\kappa(V'(\xi)+2B_0V'(\xi))}\bigg[\kappa\partial_i (B_iV'(\xi))+2\kappa B_0 B_i V''(\xi)[\pi^i+\partial_i B_0]-\partial_\mu J^\mu \bigg], \\  
&& \dot{\Pi}_0=\partial_j\Pi^j-\kappa B_0V'(\xi)-J^0.
\label{Bzero}
\end{eqnarray}

Up to here, we can numerate significant differences between the bumblebee model and standard electrodynamics: i) the number of d.o.f. is different, ii) the equations of motion do not match and iii) there is no gauge invariance nor current conservation in the bumblebee model.

Naively, under the condition $(B _\mu ( \mathbf{x},t) B^\mu(\mathbf{x},t)-b^2)=0$, which implies $V(\xi(\mathbf{x},t))=V(0)=0$ and $V ^{\prime}  (\xi(\mathbf{x},t))=V ^{\prime}(0)=0$, it can be seen that i) the canonical Hamiltonian density in Eq. (\ref{hamiltoniano1establ}) is positive definite and turns out to be the corresponding canonical Hamiltonian density of standard electrodynamics with the Gauss law strongly equal to zero, ii) the constraint $\phi_2$ in Eq. (\ref{constricciones}) becomes the standard Gauss law and iii) the time evolution of $B_j$ and $\Pi^j$ in Eq. (\ref{equmov1}) is the same as that one in ED. The authors in Ref. \cite{Potting1} noted these particularities and, using the potential $V\sim (B_\mu B^\mu-b^2)^2$, analyzed at classical level the bumblebee model and proved that under the initial conditions $(B_\mu B^\mu-b^2)|_{t_{0}}=0$ and $\partial_\mu J^\mu|_{t_{0}}=0$ the quantity $(B_\mu(\mathbf{x},t) B^\mu(\mathbf{x},t)-b^2)$ remains zero all the time. They concluded that if the phase space is restricted to those solutions satisfying the initial condition $(B_\mu B^\mu-b^2)|_{t_{0}}=0$ together with current conservation, then the bumblebee model reproduces standard electrodynamics in a non-linear gauge $(A_\mu A^\mu-b^2)=0$. However, the equivalence between bumblebee models and ED is more subtle, and here we fill in this gap.

Our main goal here is to work on some issues about this statement and we prove it for more general potentials and in a more formal way. In particular, we clarify the following: a) At classical level, in Ref. \cite{Potting1} was not discussed standard electrodynamics in a non-linear gauge $(A_\mu A^\mu-b^2)=0$. Fixing the gauge in the Hamiltonian approach, following Dirac's method, means to impose as many suitable gauge constraints ``by hand'' as there are first-class constraints (in this case we have two); these gauge constraints have to be admissible and convert the first-class constraints into second-class constraints, and then we can introduce Dirac brackets to study the dynamics of the model. In such a way, to prove the equivalence between the bumblebee model and standard electrodynamics, first it is necessary to study the electrodynamics in the non-linear gauge $(A_\mu A^\mu-b^2)=0$ by finding the corresponding Dirac brackets, and then verify the compatibility of this procedure with the bumblebee model when the initial condition $(B_\mu B^\mu-b^2)=0$ is chosen. b) At quantum level the equivalence has not been explored and requires to make clear some points. Fixing the gauge in any gauge theory requires the introduction of ghost particles (via the BRST method \cite{BRST}, for example), which play a fundamental role as internal particles when physical processes are calculated.  Therefore, to prove the proposed equivalence one should study their contributions. In this case, a possible decoupling of such ghosts is by no means clear, especially due to the non-linear character of the proposed gauge $A_\mu A^\mu-b^2=0$. If the ghosts have contributions to physical processes in standard electrodynamics when the non-linear gauge $A_\mu A^\mu-b^2=0$ is used, then the equivalence with the bumblebee model cannot be accomplished. Since the bumblebee model is not gauge invariant, ghosts contributions arising from the gauge fixing process will not be present. Also, for the bumblebee model, it should be analyzed how to introduce the condition $(B_\mu B^\mu-b^2)=0$ on the solutions in a consistent manner. Likewise, the contributions arising from the potential $V(\xi)$ have to be studied (this is not present in standard electrodynamics).


\section{Equivalence at classical level}
\label{parteclasica}

Before embarking into the bumblebee model, we first discuss the Hamiltonian formulation of standard electrodynamics in the non-linear gauge $A_\mu A^\mu-b^2=0$. The Lagrange density of standard electrodynamics is given by
\begin{equation}
\mathcal{L}_{ED}(A_\mu)=-\frac{1}{4}F_{\mu\nu}F^{\mu\nu}-A_\mu J^\mu,
\label{EDLdensity}
\end{equation}
where $F_{\mu\nu}=\partial_\mu A_\nu-\partial_\nu A_\mu$. The canonical momenta are 
\begin{equation}
\frac{\partial \mathcal{L}_{ED}}{\partial\dot{A}^0}=\Pi^0_{ED}=0,\quad\quad\quad\frac{\partial \mathcal{L}_{ED}}{\partial\dot{A}_j}=\Pi^j_{ED}=\partial_0 A_j- \partial_j A_0,
\end{equation}
which satisfy the non-zero Poisson brackets
\begin{equation}
\{A_\mu(\mathbf{x}),\Pi_{ED}^\nu(\mathbf{y})\}=\delta_\mu\,^\nu \delta(\mathbf{x}-\mathbf{y}).
\end{equation}
The canonical Hamiltonian density can be written as
\begin{equation}
\mathcal{H}^{ED}_c=\frac{1}{2}(\Pi_{ED}^j)^2+\frac{1}{4}(F_{jk})^2-A_0(\partial_i \Pi_{ED}^i-J^0)+A_i J^i.
\end{equation}
Two first-class constraints (FCC) are present
\begin{equation}
\tilde{\phi}_1=\Pi^0_{ED},\quad\quad\quad\tilde{\phi}_2=\partial_i \Pi^i_{ED}-J_0.
\end{equation}
Applying Dirac's algorithm, the number of d.o.f. is two. The corresponding gauge transformations generated from the first class constraints (FCC) are $\delta A_\mu(x)=\{A_\mu(x),\int (-\dot{\Lambda}(y) \tilde{\phi}_1(y)+\Lambda(y)\tilde{\phi}_2(y))dy\}=-\partial_\mu \Lambda(x)$. 

FCC imply unphysical degrees of freedom, which must be removed to obtain the reduced phase space. To this end we introduce Dirac brackets after imposing the gauge fixing constraints. In our case, one of the gauge fixing constraints is $\tilde{\phi}_3=A_\mu A^\mu-b^2$. This has to be complemented with another suitable gauge fixing constraint. We choose $\tilde{\phi}_4=A_0$. It can be verified that the set of constraints $\{\tilde{\phi}_i, i=1,2,3,4\}$ is second-class.

In order to obtain the corresponding Dirac brackets $\{\mathcal{A},\mathcal{B}\}_D$, we require the matrix constructed with the Poisson brackets $\{\mathcal{A},\mathcal{B}\}$ of the constraints $\tilde{\phi}_i$, which is given by

\begin{equation}
\label{Diracbra1}
M_{ij}(\mathbf{x},\mathbf{y})=\{\tilde{\phi}_i(\mathbf{x}),\tilde{\phi}_j(\mathbf{y})\}=
\left( {\begin{array}{cccc}
   0 & 0 & -2 A_0 & -1 \\      
   0 & 0 & -2\partial_i A^i &0 \\
   2A_0 & 2\partial_i A^i & 0 &0 \\
   1 & 0 & 0 &0 \\     
   \end{array} } \right) \delta(\mathbf{x}-\mathbf{y}).
\end{equation}
The inverse matrix $(M^{-1})_{ij}$, such that $\int d^3z\, M_{il}(\mathbf{x},\mathbf{z}) (M^{-1})_{lj}(\mathbf{z},\mathbf{y})=\delta_{ij}\delta(\mathbf{x}-\mathbf{y}) $ is
\begin{equation}
\label{Diracbra2}
(M^{-1})_{ij}(\mathbf{x},\mathbf{y})=
\left( {\begin{array}{cccc}
   0 & 0 & 0 & 1 \\      
   0 & 0 & \frac{1}{2\partial_i A^i} & -\frac{A_0}{\partial_i A^i} \\
   0 & -\frac{1}{2\partial_i A^i} & 0 &0 \\
   -1 & \frac{A_0}{2\partial_i A^i} & 0 &0 \\     
   \end{array} } \right) \delta(\mathbf{x}-\mathbf{y}).
\end{equation}
Dirac brackets are defined as follows
\begin{equation}
\label{Diracbra3}
\{\mathcal{A}(\mathbf{x}),\mathcal{B}(\mathbf{y})\}_D=\{\mathcal{A}(\mathbf{x}),\mathcal{B}(\mathbf{x})\}-\int\, d^3u\, d^3v\{\mathcal{A}(\mathbf{x}),\Phi_i(\mathbf{u})\}(M^{-1})^{ij}\{\Phi_j(\mathbf{v}),\mathcal{B}(\mathbf{y})\},
\end{equation}
where $\Phi_i$ denote any of the constraints $\tilde{\phi}_i$, with $i=1,2,3,4$. The above allows us to remove the variables $(A_0,\Pi^0_{ED})$ and work with the algebra for the remaining variables $(A_i,\Pi^j_{ED})$, which is given by
\begin{eqnarray}
\label{Dalgebra2}
&&\{A_i(\mathbf{x}),A_j(\mathbf{y})\}_D=0,\quad\quad\quad \\ \nonumber
&&\{\Pi_{ED}^i(\mathbf{x}),\Pi_{ED}^j(\mathbf{y})\}_D=0,\quad\quad\quad\\ \nonumber
&&\{A_i(\mathbf{x}),\Pi_{ED}^j(\mathbf{y})\}_D=\delta_i\,^j\delta(\mathbf{x}-\mathbf{y})-\frac{A_j(\mathbf{y})}{\partial_l A^l(\mathbf{y})}\partial_{yi}\delta(\mathbf{x}-\mathbf{y}).
\end{eqnarray}
We can set strongly equal to zero the constraints $\tilde{\phi}_i$ and describe the dynamics through Dirac brackets with the Hamiltonian density  
\begin{equation}
\mathcal{H}_F=\frac{1}{2}(\Pi_{ED}^j)^2+\frac{1}{4}(F_{jk})^2+A_i J^i.
\label{Hamil2}
\end{equation}
At this point the Hamiltonian (\ref{Hamil2}) together with the algebra (\ref{Dalgebra2}) are sufficient to determine the dynamics of standard electrodynamics in the non-linear gauge $A_\mu A^\mu-b^2=0$. The time evolution for any quantity can be obtained from them as
\begin{equation}
\dot{\mathcal{Q}}(A_i,\Pi^j_{ED})=\{\mathcal{Q},\mathcal{H}_F\}_D+\frac{\partial \mathcal{Q}}{\partial t}.
\end{equation}
To establish an equivalence at classical level between the bumblebee model and standard electrodynamics in the non-linear gauge $A_\mu A^\mu-b^2=0$, we have to prove that:  i) from the bumblebee model it is possible to construct an algebra with brackets alike to those in Eq. (\ref{Dalgebra2}) and ii) obtain the same time evolution when the corresponding algebra and bumblebee Hamiltonian $\mathcal{H}^B$, plus suitable initial conditions, are used.

Let us explore the time evolution of the quantity $(B_\mu B^\mu-b^2)$ under the dynamics of the bumblebee model. From Eqs. (\ref{equmov1})-(\ref{Bzero}) we obtain
\begin{equation}
\partial_0(B_0^2-B_i^2-b^2)=\frac{1}{\kappa(V'(\xi)+2B_0^2V''(\xi))}\bigg[2\kappa B_0\partial_i[B_iV'(\xi)]-2B_0 \partial_\mu J^\mu-2\kappa V'(\xi)B_j(\Pi^j+\partial_j B_0) \bigg].
\end{equation}
This equation together with Eq. (\ref{Bzero}) reveal that if the initial conditions $B_0|_{t_{0}}=0$, $(B_\mu B^\mu-b^2)|_{t_{0}}=0$ and $\partial_\mu J^\mu |_{t_{0}}=0 $ are chosen, then the relations $(B_\mu B^\mu-b^2)(t)=0$ and $B_0(t)=0$ remain valid for all time. Therefore, with these particular initial conditions, the dynamics of the bumblebee model can be restricted, without inconsistencies or extra conditions, to the set of constraints $\{\phi_1=\Pi^0,\phi_2=\partial_j\Pi^j-\kappa B_0V'(\xi)-J^0,\phi_3=B_\mu B^\mu-b^2,\phi_4=B_0\}$. Using $\phi_3$ and $\phi_2$ together with $V'(\xi=0)=0$, the set of constraints $\{\phi_i\}$ can be rewritten as the set $\{\Lambda_i\}$, composed of
\begin{equation}
\Lambda_1=\Pi^0=0,\quad\quad\Lambda_2=\partial_j\Pi^j-J^0=0,\quad\quad \Lambda_3=B_\mu B^\mu-b^2=0,\quad\quad\Lambda_4=B_0=0.
\end{equation}
Note that they have the same structure as those that appear in standard electrodynamics in the non-linear gauge $A_\mu A^\mu-b^2=0$. We can use Dirac brackets to define an algebra in which the constrains $\Lambda_i$ are preserved. This ensures that the dynamics does not leave the constraint surface. The construction of these brackets proceeds as the same way as in Eqs. (\ref{Diracbra1})-(\ref{Diracbra3}). Consequently, the brackets in Eq. (\ref{Dalgebra2}) describe the algebra of the bumblebee model with initial conditions $B_0|_{t_{0}}=0$, $(B_\mu B^\mu-b^2)|_{t_{0}}=0$ and $\partial_\mu J^\mu |_{t_{0}}=0$. Certainly all constraints, under Dirac brackets, satisfy $\{\Lambda_i,\mathcal{B}\}_D=0$ to any quantity $\mathcal{B}$; in particular, $\dot{\Lambda}_i=\{\Lambda_i,H^B\}_D=0$.

We have to point out that in the last procedure the fundamental property is that, under the aforementioned initial conditions, the relations $B_0=0$ and $(B_\mu B^\mu-b^2)=0$ remain valid for all time, which are derived from the dynamics of the bumblebee model. If this property is not fulfilled, then Dirac brackets are not well defined, because in this case we would need more constraints to ensure that the dynamics does not leave the constraint surface given by the set $\{\Lambda_i\}$. Dirac brackets constructed with more constraints than $\{\Lambda_i\}$  would not correspond with those in Eq. (\ref{Dalgebra2}). In other words, we have started on a surface of the phase space where the bumblebee potential does not have contributions and we have proved that the dynamics of the bumblebee model by itself evolves just on this surface, without having to impose extra conditions.

After introducing Dirac brackets together with the imposition of constraints strongly zero $(\Lambda_i=0)$, the bumblebee Hamiltonian density $H^B_c$ in Eq. (\ref{hamiltoniano1establ}) reduces to the Hamiltonian density $H_F$ of electrodynamics in Eq. (\ref{Hamil2}).

In this way, the dynamics of the bumblebee model with initial conditions $B_0|_{t_{0}}=0$, $(B_\mu B^\mu-b^2)|_{t_{0}}=0$ and $\partial_\mu J^\mu |_{t_{0}}=0$, which is derived from the algebra in Eq. (\ref{Dalgebra2}) and Hamiltonian in Eq. (\ref{Hamil2}), corresponds to the dynamics of standard electrodynamics in the non-linear gauge $A_\mu A^\mu-b^2$ \footnote{A different option to introduce the condition $B_\mu B^\mu-b^2=0$ can be by means of the term $\lambda(B_\mu B^\mu-b^2)$ in the bumblebee Lagrange density, with $\lambda$ being a Lagrange multiplier. The equation of motion for $\lambda$ gives the condition $(B_\mu B^\mu-b^2)=0$ in a natural way; however, to establish the equivalence with the standard electrodynamics we would need to restrict the model to values with $\lambda=0$, which can be done by calculating its time evolution and taking the suitable initial conditions. Namely, with the introduction of the Lagrange multiplier is not necessary to choose the initial condition $B_\mu B^\mu-b^2=0$ but  requires the initial condition $\lambda=0$. It is in this sense that there is not an improvement for our purposes if we introduce the Lagrange multiplier.}.

Notice that the subsequent emergence of standard electrodynamics, after imposing the initial conditions, guarantees current conservation for all times, as a consequence of the equations of motion.


\section{Equivalence at quantum level}
\label{partecuantica}

In this section we study from a quantum perturbative perspective both standard electrodynamics and the bumblebee model.  The aim is to compare the Feynman rules and propagators of these theories.

\subsection{The quantum approach of standard electrodynamics in the non-linear gauge $A_\mu A^\mu-b^2$}
\label{Quantumelectro}

The quantum analysis of standard electrodynamics in the non-linear gauge  $A_\mu A^\mu-b^2$ was recently studied by one of us in Ref. \cite{Nambuphotons}. In the following we summarize the main results.

In this case, the BRST quantization formalism \cite{BRST} provides an appropriate method to address the gauge fixing process in standard electrodynamics. To this end, the fermionic nilpotent transformation $\tilde{\delta}$ is introduced, together with the fields $c$, $\bar{c}$ and $\check{b}$, satisfying 
\begin{equation}
\tilde{\delta} A_\mu=\partial_\mu c,\quad \tilde{\delta}c=0,\quad\tilde{\delta}\bar{c}=i\check{b},\quad\tilde{\delta}\check{b}=0.
\end{equation}
The BRST invariant Lagrange density is written as
\begin{eqnarray}
\mathcal{L}_{ED}^{BRST}(A_\mu)=-\frac{1}{4}F_{\mu\nu}F^{\mu\nu}+A_\mu J^\mu+i\tilde{\delta}\bigg[\bar{c}\bigg((A_\mu A^\mu-b^2)+\frac{\check{b}}{2\alpha}\bigg)\bigg],
\end{eqnarray}
where the gauge fixing condition $(A_\mu A^\mu-b^2)=0$ is explicitly introduced. After the $\tilde{\delta}$ variation and eliminating $\check{b}$ from its equation of motion, the Lagrange density becomes 
\begin{equation}
\mathcal{L}_{ED}^{BRST}(A_\mu)=-\frac{1}{4}F_{\mu\nu}F^{\mu\nu}+A_\mu J^\mu-\frac{\alpha}{2}(A_\mu A^\mu-b^2)^2-2A^\mu i \bar{c}\partial c,
\label{EDBRST}
\end{equation}
which plainly shows the contributions from the fixing term $\alpha(A_\mu A^\mu-b^2)^2/2$ and the Faddeev-Popov ghosts $c$ and $\bar{c}$. 

At this stage it is convenient to employ the parametrization proposed in Ref. \cite{parametri12}
\begin{equation}
A_\mu(a_\rho)=a_\mu+n_\mu(b^2- a^2)^{1/2},
\label{parametrization1}
\end{equation}   
where $n_\mu$ is a unit timelike vector. An important consequence of this field redefinition is the relation 
\begin{equation}
(A_\mu A^\mu-b^2)^2=4(n\cdot a)^2(b^2-a^2).
\label{relation1}
\end{equation}
When the parametrization in Eq. (\ref{parametrization1}) is substituted into the Lagrange density $\mathcal{L}_{ED}^{BRST}(A_\mu)$ in Eq. (\ref{EDBRST}) we obtain a complicated non-linear expression $\mathcal{L}_{ED}^{BRST}(A_\mu(a_\rho))$ in terms of the $a_\mu$ field; however, this one still corresponds to the Lagrange density of standard electrodynamics, written in a very unconventional form. For example, the electromagnetic stress tensor becomes
\begin{equation}
F_{\mu\nu}=f_{\mu\nu}+(n_\nu\partial_\mu-n_\mu \partial_\nu)(b^2-a^2)^{1/2},
\end{equation}
where
\begin{equation}
f_{\mu\nu}=\partial_\mu a_\nu-\partial_\nu a_\mu.
\end{equation}
In terms of the $a_\mu$ field we rewrite Eq. (\ref{EDBRST}) as 
\begin{equation}
\mathcal{L}_{ED}^{BRST}(a_\mu)= \mathcal{L}_{ED}(A_\mu(a_\rho))-2\alpha b^2(n\cdot a)^2+2\alpha a^2(n\cdot a)^2-2\bigg(a^\mu+n^\mu(b^2-a^2)^{1/2} \bigg)i\bar{c}\partial_\mu c,
\label{lagranBRST34}
\end{equation}
where $\mathcal{L}_{ED}(A_\mu(a_\rho))$ corresponds to the standard Lagrange density of electrodynamics in Eq. (\ref{EDLdensity}) rewritten in terms of the $a_\mu$ field. Expanding $\mathcal{L}_{ED}(A_\mu(a_\rho))$ in powers of $(a ^{2} / b ^{2}) < 1$ we get
\begin{eqnarray}
\nonumber
\mathcal{L}_{ED}(A_\mu(a_\rho))&=&-\frac{1}{4} f_{\mu\nu}f^{\mu\nu}-\frac{b}{2}f_{\mu\nu}(n^\mu\partial^\nu-n^\nu\partial^\nu)\bigg[\sum_{l=1}^\infty
\begin{pmatrix} 1/2 \\ l \end{pmatrix} \bigg(\frac{a^2}{b^2}\bigg)^l\bigg] \\ \nonumber
&&-\frac{b^2}{4}
(n_\mu\partial_\nu-n_\nu\partial_\nu)\bigg[\sum_{l=1}^\infty
\begin{pmatrix} 1/2 \\ l \end{pmatrix} \bigg( \frac{a^2}{b^2}\bigg)^l
 \bigg]
 (n^\mu\partial^\nu-n^\nu\partial^\nu)\bigg[\sum_{m=1}^\infty
\begin{pmatrix} 1/2 \\ m \end{pmatrix}  \bigg(\frac{a^2}{b^2}\bigg)^m\bigg] \\
&&- a_\mu J^\mu+ b \,n_\mu J^\mu \bigg[\sum_{l=0}^\infty
\begin{pmatrix} 1/2 \\ l \end{pmatrix}  \bigg(\frac{a^2}{b^2}\bigg)^l
 \bigg],
 \label{Lagranfreeexp}
\end{eqnarray}
where in the first three summations we have started the serie from $l=1$ due that the contribution from $l=0$ produces a constant term, which vanishes with the partial derivate in the brackets. This shows that the only quadratic terms in $\mathcal{L}_{ED}^{BRST}(a_\mu)$, that are not coupled to the ghost or the external current, are given by
\begin{equation}
\mathcal{L}_{ED_{quad}}^{BRST}(a_\mu)=-\frac{1}{4} f_{\mu\nu}f^{\mu\nu}-2\alpha b^2 (n\cdot a)^2.
\label{quadractict}
\end{equation}

The remaining terms are cubic and higher orders in powers of the field $a^\mu$. Notice that the condition $-2\alpha b^2 (n\cdot a)^2$ coincides to the choice of an axial gauge in standard electrodynamics, while that $2\alpha a^2(n\cdot a)^2$ and the higher order terms in Eq. (\ref{Lagranfreeexp}) can be interpreted as extra Yang-Mills-type interactions. Namely, the parameterization in Eq. (\ref{parametrization1}) allows to interpret the $a_\mu$ field as photons in the gauge $(n\cdot a)=0$ with non-linear interactions. The propagator in the axial gauge, which arises from the quadratic terms in Eq. (\ref{quadractict}), can read off from Ref. \cite{Nambuphotons,axialgauge}
\begin{equation}
D^{ED}_{\mu\nu}(k)=\frac{-i}{k^2+i\epsilon}\bigg[\eta_{\mu\nu}-\frac{k_\mu n_\nu+n_\mu k_\nu}{(n\cdot k)}+k_\mu k_\nu \frac{n^2+\frac{k^2}{4\alpha b^2}}{(n\cdot k)^2} \bigg],
\end{equation}
satisfying 
\begin{equation}
k^\mu D_{\mu\nu}^{ED}(k)=0,\quad \quad n^\mu D_{\mu\nu}^{ED}(k)=\frac{-i}{k^2+i\epsilon}\bigg[ \frac{k^2 k_\nu}{4b^2(n\cdot k)} \bigg]\frac{1}{\alpha}.
\end{equation}
Now we take the so-called pure (homogeneous) axial gauge, defined by $\alpha\rightarrow\infty$. In this gauge the above expressions reduce to
\begin{equation}
D^{ED}_{\mu\nu}(k)=\frac{-i}{k^2+i\epsilon}\bigg[\eta_{\mu\nu}-\frac{k_\mu n_\nu+n_\mu k_\nu}{(n\cdot k)}+k_\mu k_\nu \frac{n^2}{(n\cdot k)^2} \bigg],\quad k^\mu D_{\mu\nu}^{ED}(k)=0,\quad \quad n^\mu D_{\mu\nu}^{ED}(k)=0.
\label{propagatorED1}
\end{equation}
Next we deal with the contributions to physical processes arising from the fixing terms $-2\alpha b^2(n\cdot a)^2$ and $2\alpha a^2(n\cdot a)^2$ in Eq. (\ref{lagranBRST34}), which do not depend on the Faddeev-Popov ghosts and are not present in the bumblebee model. The term $a^2(n\cdot a)^2$ produces a four-photon vertex $V_{\alpha\beta\mu\nu}$,

\begin{equation}
V_{\alpha\beta\mu\nu}\sim \alpha (\eta_{\alpha\beta}n_\mu n_\nu+\textrm{perm}).
\label{vertex1}
\end{equation}

When we attach external on-shell photons to $V_{\alpha\beta\mu\nu}$, via the corresponding propagators, the conditions $n^\mu D^{ED}_{\mu\nu}(k)=0=k^\mu D^{ED}_{\mu\nu}(k)$ lead to a zero contribution to any physical process. Also, these conditions imply that their polarization vectors $\epsilon_\mu(k)$ must satisfy $n^\mu\epsilon_\mu(k)=0$ together with the transversality condition $k^\mu\epsilon_\mu(k)=0$ \cite{nambu,Nambuphotons}. In turn, for internal photon lines attached to $V_{\alpha\beta\mu\nu}$ the vertex has two contributions of the type $n^\mu D_{\mu\nu}$, producing a factor of $1/\alpha^2$, with a net result going like $1/\alpha$, which is zero in the limit $\alpha\rightarrow\infty$. The same argument can be applied to the gauge fixing term proporcional to $(n\cdot a)^2$. In this way, the terms involving $(n\cdot a)$ just implement the gauge condition $(n\cdot a)=0$.

Consequently the gauge-fixed Lagrange density $\mathcal{L}_{ED}^{BRST}(a_\mu)$ for electrodynamics, in the pure axial gauge, reads 
\begin{equation}
\mathcal{L}_{ED}^{BRST}(a_\mu)=\mathcal{L}_{ED}(A_\mu(a_\rho))+\mathcal{L}_{GHOST},
\end{equation}
where $\mathcal{L}_{GHOST}$ corresponds to the contribution involving the Faddeev-Popov ghosts. One of the main results in Ref. \cite{Nambuphotons} is precisely to prove that, using the parametrization (\ref{parametrization1}) and to any physical process, the ghosts decouple in the non-linear gauge $A_\mu A^\mu-b^2=0$, or equivalently in the axial gauge $(n\cdot a)=0$ plus added non-linear interactions; therefore, there are not contributions arising from the $\mathcal{L}_{GHOST}$.

In this way, the contributions to Feynman amplitudes to any physical process of the gauge-fixed Lagrange density $\mathcal{L}_{ED}^{BRST}(a_\mu)$ in Eq. (\ref{lagranBRST34}) are constructed only from $\mathcal{L}_{ED}(A_\mu(a_\rho))$. The corresponding Feynman rules for $\mathcal{L}_{ED}(A_\mu(a_\rho))$ are given in Ref. \cite{parametri12}. Since the free part of the bumblebee model (\ref{LagrangianBumblebee}) can be seen as $\mathcal{L}_{ED}(A_\mu)$ supplemented with a potential of form (\ref{potential12}), the Feynman rules arising from $\mathcal{L}_{ED}(A_\mu)$ will also emerge in the bumblebee model.


\subsection{The quantum approach of bumblebee model}

Now we switch to the bumblebee model. We are interested in studying its  quantum equivalence with the fixed-gauge standard electrodynamics analyzed in Section \ref{Quantumelectro}. To this end, we focus on the bumblebee model with solutions restricted to the condition $B_\mu B^\mu-b^2=0$. Regarding the quantization of the unrestricted bumblebee model, the introduction of the aforementioned condition allows some simplifications in the quantum approach. 

The quantization of the bumblebee model was performed in Ref. \cite{HERNASKI} following the Stueckelberg method \cite{stueckel}, which consists of the introduction of a local symmetry in the Lagrangian density by the enlargement of the field content to turn second-class constraints to first-class. The quantization of the resulting theory with first-class constraints is achieved following the standard methods of gauge theories, such as Faddeev-Popov method \cite{popov} or BRST method \cite{BRST}. 

Even though the Stueckelberg method allows to study the quantization of the bumblebee model, and other properties as stability, the possible equivalence with standard electrodynamics is not completely clear under this scheme. The introduction of auxiliar fields, due to the Stueckelberg method, plus the perturbative approach employed do not allow to establish a direct connection with standard electrodynamics.

In the present work we apply a different process. The strategy we follow is the employment of the parametrization in Eq. (\ref{parametrization1}) and obtaining the ingredients for the construction of Feynman amplitudes, i.e. Feynman rules and propagator. Then we compare these with those stemming from the fixed-gauge electrodynamics.

We start from the Lagrange density of the bumblebee model in Eq. (\ref{LagrangianBumblebee}). With the replacement  $A_\mu\rightarrow B_\mu$ in the parametrization (\ref{parametrization1}) we can rewrite the bumblebee Lagrange density as
\begin{equation}
\mathcal{L}_B(a_\mu)= \mathcal{L}_{B_{f}}(a_\mu)-\frac{\kappa}{2}V(2(n\cdot a)(b^2-a^2)^{\frac{1}{2}}),
\label{densityLB}
\end{equation} 
where the first term $ \mathcal{L}_{B_{f}}(a_\mu)$ corresponds to the free part $-\frac{1}{4} B_{\mu\nu} B^{\mu\nu}+B_\mu J^\mu$  and the second term to the potential $V=V(B_\mu B^\mu-b^2)$, both rewritten in terms of the $a_\mu$ field. To make contact between the parametrization in Eq. (\ref{parametrization1}) and the nonzero vacuum expectation value of the bumblebee model, we can express the VEV, $\langle B_\mu\rangle=\mathcal{B}_\mu$, in Eq. (\ref{vacio1}) as 
\begin{equation}
\mathcal{B}_\mu=n_\mu b.
\end{equation}
Under this description $n_\mu$ defines the direction of the vacuum, while $b$ its magnitude. Notice that the condition $\mathcal{B}_\mu \mathcal{B}^\mu=b^2$ is automatically satisfied.

Next we address the incorporation of the condition $B_\mu B^\mu-b^2=0$. According to Eq. (\ref{relation1}), this condition can be translated to 
\begin{equation}
(n\cdot a)=0,
\label{conditionna}
\end{equation}
when the field redefinition (\ref{parametrization1}) is employed. The $a_\mu$ fields, satisfying the condition in Eq. (\ref{conditionna}), define the three d.o.f. of the bumblebee model and they are orthogonal to the vacuum direction $n^\mu$, so that they describe the Goldstone bosons of the model. 

The propagator satisfying the condition (\ref{conditionna}) can be constructed from the bumblebee Lagrange density with solutions restricted to $(n\cdot a)=0$, which we write as 
\begin{equation}
\mathcal{L}_B |_{((n\cdot a)=0)}= \mathcal{L}_B(a_\mu)+\lambda(n_\mu a^\mu),
\end{equation}
where $\lambda$ is a Lagrange multiplier and $\mathcal{L}_B(a_\mu)$ is the Lagrange density in Eq. (\ref{densityLB}). The additional contribution with the Lagrange multiplier cannot be interpreted as a gauge fixing term (there is no gauge to fix in the bumblebee model) and only implements the condition $(B_\mu B^\mu- b^2)=0$, or $n_\mu a^\mu=0$ in terms of the $a_\mu$ fields, in a covariant way. Note that we do not make $(B_\mu B^\mu- b^2)$ strongly equal to zero at the level of the Lagrange density, which would mean to consider zero the bumblebee potential. Instead we include the condition over the solutions of the model through the Lagrange multiplier.

To go further we consider potentials as in Eq. (\ref{potential12}) and identify the quadratic terms in the bumblebee Lagrange density. The relation
\begin{equation}
\frac{(B_\mu B^\mu-b^2)^m}{m}=\frac{2^m(n\cdot a)^m(b^2-a^2)^{\frac{m}{2}}}{m}
\end{equation}
allows to classify the cases: (i) $m=2$ and (ii) $m>2$. For the case with $m=2$, the propagator is  defined by the terms
\begin{equation}
\mathcal{L}_{\rm KIN}^{B}\mathcal{=-}\frac{1}{4}f_{\mu \nu }f^{\mu \nu
}+\kappa b^2(n\cdot a)^2+\lambda \left( n\cdot a \right) -J^{\mu }a_{\mu },
\end{equation}
leading to the equations of motion
\begin{equation}
\partial ^{\mu }f_{\mu \nu }+(\lambda+\kappa b^2(n\cdot a) )n_{\nu }=J_{\nu },\quad\quad n\cdot a=0.  
\label{EQMOTANM}
\end{equation}
After solving for the quantity $(\lambda+\kappa b^2(n\cdot a) )$ by multiplying the first equation in (\ref{EQMOTANM}) by $n^{\nu }$ and inserting back the result, we get
\begin{equation}
\left[ \partial ^{2}\eta ^{\nu \gamma }-\left( \delta _{\;\alpha }^{\nu }-
\frac{n^{\nu }n_{\alpha }}{n^{2}}\right) \partial ^{\alpha }\partial
^{\gamma }\right] a_{\gamma }\equiv O^{\nu \gamma }a_{\gamma }=\left( \delta
_{\alpha }^{\nu }-\frac{n^{\nu }n_{\alpha }}{n^{2}}\right) J^{\alpha }=
\tilde{J}^{\nu },
\end{equation}
which defines the operator $O^{\nu \gamma }$ to be inverted. Since we have $n_{\nu }O^{\nu \gamma }a_{\gamma }=0$, we must find the bumblebee propagator $D_{\gamma \rho }^B$ in the subspace orthogonal to $n_{\nu }$ by demanding
\begin{equation}
O^{\nu \gamma }D_{\gamma \rho }^B=\left( \delta _{\rho }^{\nu }-\frac{n^{\nu
}n_{\rho }}{n^{2}}\right),  
\label{CONDPROP}
\end{equation}
where the LHS is the unit in that subspace. Writing the most general form of $D_{\gamma \rho }^B$ as
\begin{equation}
D_{\gamma \rho }^B(k)=-\frac{i}{k^{2}+i\epsilon }\left[ \eta _{\gamma \rho }+
\frac{A}{\left( n\cdot k\right) }\left( n_{\gamma }k_{\rho }+k_{\gamma
}n_{\rho }\right) +\frac{B}{k^{2}}k_{\gamma }k_{\rho }+C\frac{n_{\gamma
}n_{\rho }}{n^{2}}\right]
\end{equation}
and imposing the conditions\ (\ref{CONDPROP}) we obtain
\begin{equation}
D_{\gamma \rho }^B(k)=-\frac{i}{k^{2}+i\epsilon }\left[ \eta _{\gamma \rho }-
\frac{n_{\gamma }k_{\rho }+k_{\gamma }n_{\rho }}{\left( n\cdot k\right) }+
\frac{n^{2}k_{\gamma }k_{\rho }}{\left( n\cdot k\right) ^{2}}\right],  
\label{FOTPROPANM}
\end{equation}
which naturally incorporates the condition $n^{\gamma }D_{\gamma \rho }^B=0$. Transversality on $k_\mu$ on-shell vectors, $k^\gamma D_{\gamma \rho }^B(k)=0$, is also fulfilled. 

In the case (ii) with $m>2$ the potential does not contribute to the calculation to the propagator; however, the above process can be applied again, giving the same propagator as in Eq. (\ref{FOTPROPANM}). Notice that the propagator $D_{\gamma \rho }^B(k)$ is the same that appears in Eq. (\ref{propagatorED1}) for the fixed-gauge electrodynamics. Thus, the condition $(n\cdot a)=0$ is implemented in the same way for each model in terms of the propagator. The requirement $(n\cdot a)=0$ holds in both cases; however, it is for different reasons: it is a gauge condition in ED, while it corresponds to a restriction, introduced by a Lagrange multiplier, over the solutions of the bumblebee model. The transversality condition $k^\mu \epsilon_\mu(k)=0$ on $k_\mu$ on-sell photons, derived from the propagator satisfying $k^\gamma D_{\gamma \rho }^B(k)=0$, implements the Gauss' law in the bumblebee model \textit{\`{a} la} Dirac upon the initial states. The current $J^\mu$ is conserved, as it is a Noether current arising from the gauge invariance derived from the Gauss' law, which is only valid on the states with the condition $(n\cdot a)=0$.   

Since the interaction terms $\kappa(n\cdot a)^n(b^2-a^2)^{\frac{n}{2}}$ in Eq. (\ref{densityLB}) produce the vertexes $\tilde{V}^{1n}_{\mu\nu}\sim \kappa (n_\mu n_\nu)^n$, $\tilde{V}^{2n}_{\mu\nu\alpha\beta}\sim \kappa (\eta_{\mu\nu}n_\alpha n_\beta+ \textrm{perm}.)^{\frac{n}{2}}$, which are similar to $V_{\mu\nu\alpha\beta}$ in Eq. (\ref{vertex1}), the contributions to Feynman amplitudes from the potential in the bumblebee model cancel out due to the fact that the conditions $n^{\gamma }D_{\gamma \rho }^B=0$ and $k^\gamma D_{\gamma \rho }^B(k)=0$ are satisfied.  In this way, the Feynman rules for the bumblebee model, with solutions restricted to $(n\cdot a=0)$, arise only from the free part $\mathcal{L}_{B_{f}}(a_\mu)$. Let us empathize that the $\mathcal{L}_{B_{f}}(a_\mu)$ in Eq. (\ref{densityLB}) is the same Lagrange density $\mathcal{L}_{ED}(A_\mu(a_\rho))$ that defines the fixed-gauge electrodynamics in Eq. (\ref{lagranBRST34}); therefore, Feynman rules, propagators and interactions of the bumblebee model, restricted to the condition $(n\cdot a=0)$, correspond with those in the fixed-gauge standard electrodynamics.


\section{Summary}
\label{conclusiones}

Since many approaches of quantum gravity suggest or have processes in which Lorentz invariance is broken, Lorentz violation (LV) became an active and rich line of research \cite{Others,string,brane,noncumative,emergentG,DSR,aether,topology,cosmology,gaugeb}. Among the mechanisms to introduce LV in our current theories, spontaneous symmetry breaking is an appropriate approach to break Lorentz invariance. For example, the Standard Model Extension (SME) \cite{SME1}, the most used framework to parametrize deviations of Lorentz symmetry, arises from this concept. Close to the SME we find the so-called bumblebee models, which are effective field vector theories with a non-zero vacuum expectation value that spontaneously breaks Lorentz symmetry. Due to the manageability of these models, phenomenological and theoretical relevant properties of Lorentz violation and spontaneous symmetry breaking can be analyzed and disentangled.  Also, they have been proposed as an alternative to standard electrodynamics where the photons are identified as Goldstone bosons arising from the spontaneous symmetry breaking of Lorentz invariance. The above idea was investigated in Ref. \cite{Potting1} for some bumblebee models at classical level, and the key point was the restriction of the phase space to solutions whose dynamics reproduces the equations of motion of standard electrodynamics. A Hamiltonian analysis was performed in order to study such an equivalence; however, it remains some issues to fully prove it. To complete all the remaining missing pieces constitutes the main motivation of this work, particularly at quantum level and for more general potentials.

To establish an equivalence between bumblebee models and standard electrodynamics (ED), first we must find the conditions under which the effects of Lorentz violation are unobservable and, second we must verify that dynamics and calculations of physical processes are identical in both theories. In the present work, we have analyzed the bumblebee model with a potential of the form $V(B_\mu)=V(B_\mu B^\mu- b^2)$. The main result was to prove that after imposing suitable conditions, the resulting theory is equivalent to standard electrodynamics in a non-linear gauge $A_\mu A^\mu-b^2=0$, both at classical and quantum levels.

We begin by summarizing the fundamental properties of the bumblebee model and by indicating the differences with standard electrodynamics, which are: i) the number of degrees of freedom is different, ii) the equations of motion do not match and iii) there are not gauge invariance and current conservation in the bumblebee model. According to Ref. \cite{Potting1} the bumblebee model reproduces standard electrodynamics when its phase space is restricted to solutions satisfying the initial condition $B_\mu B^\mu-b^2=0$. We completed the about statement by studying how to introduce this condition both, at classical and quantum levels, in a consistent manner and by comparing the resulting model with standard electrodynamics in a non-linear gauge $A_\mu A^\mu-b^2=0$.

At classical level, we performed a Hamiltonian analysis to standard electrodynamics in the fixed gauge $A_\mu A^\mu-b^2=0$. We constructed the Dirac brackets in Eq. (\ref{Dalgebra2}) and wrote down the Hamiltonian in Eq. (\ref{Hamil2}) from which the dynamics, of ED in the fixed gauge, is derived. After that, we employed Dirac's method to study the bumblebee model. Imposing the initial conditions $B_0|_{t_{0}}=0$, $(B_\mu B^\mu-b^2)|_{t_{0}}=0$ and $\partial_\mu J^\mu |_{t_{0}}=0$ we constructed a consistent algebra, in which the condition  $B_\mu B^\mu-b^2=0$ is included and preserved for all times. This algebra and the restricted Hamiltonian to such initial conditions correspond to those stemming from the gauge-fixed standard electrodynamics in Eqs. (\ref{Dalgebra2})-(\ref{Hamil2}), respectively. Thus, the time evolution, arising from the algebra and the Hamiltonian, is the same in both theories. We also verified that there are no inconsistencies or additional constraints when the required initial conditions in the bumblebee model are imposed. 

To prove the equivalence at quantum level, the strategy followed was to prove that, after imposing the condition $B_\mu B^\mu-b^2=0$ on the solutions of the bumblebee model, the Feynman rules and propagators of standard electrodynamics in the fixed gauge $A_\mu A^\mu-b^2=0$ and those describing the bumblebee model are the same. In this way, perturbative calculations of Feynman amplitudes to any physical process in each model are indistinguishable. To this end, we begin by analyzing standard electrodynamics employing the BRST method to fix the non-linear gauge $A_\mu A^\mu-b^2=0$. After employing a very convenient parametrization $A_\mu\rightarrow a_\mu $ defined in Eq. (\ref{parametrization1}), which allows to translate the non-linear relation $A_\mu A^\mu-b^2=0$ to the expression $(n\cdot a )=0$, the Feynman rules and propagators are obtained. In the fixed-gauge Lagrange density of standard electrodynamics in Eq. (\ref{lagranBRST34}) we can identify three different kinds of contributions: a) those arising from the free part ($\frac{1}{4}F_{\mu\nu} F^{\mu\nu}+A_\mu J^\mu$ rewritten in terms of the $a_\mu$ field redefinition), b) gauge fixing terms and c) the Faddeev-Popov ghosts. Using the results of Ref. \cite{Nambuphotons}, it can be proved that the contributions from b) and c) cancel out. The above implies that the Feynman amplitudes of standard electrodynamics in the non-linear gauge are only constructed from the Feynman rules derived from the free part plus the corresponding propagator. Next we turn to the quantum description of the bumblebee model, in which we used the same parametrization in Eq. (\ref{parametrization1}). The relation $B_\mu B^\mu-b^2=0$ translated to $(n\cdot a )=0$ allows to identify the $a_\mu$ fields as the pure Goldstone bosons of the model that are orthogonal to the direction $n_\mu$ of the vacuum inducing the spontaneous Lorentz symmetry breaking. Also, it allows to work only with the three degrees of freedom of the model. The restriction over the solutions of the bumblebee model satisfying $(n\cdot a )=0$ is effectively incorporated in the calculations through the propagator $D_{\mu\nu}^B(k)$ given in Eq. (\ref{FOTPROPANM}) and satisfying $n^\mu D_{\mu\nu}^B(k)=0$ together with $k^\mu D_{\mu\nu}^B(k)=0$ for on-shell Goldstone bosons. This propagator $D_{\mu\nu}^B(k)$ is identical to the propagator $D_{\mu\nu}^{ED}(k)$ in Eq. (\ref{propagatorED1}) of standard electrodynamics in the non-linear gauge. The on-shell transversality of $D_{\mu\nu}^B(k)$ guarantees that the Gauss' law is imposed, \textit{\`{a} la} Dirac, upon the physical states \cite{nambu,Nambuphotons}. Gauge invariance, generated by the Gauss' law, is satisfied on the states with $(n\cdot a )=0$. The contributions to Feynman amplitudes arising from the bumblebee potential cancel out under the aforementioned requirements; in such a way, the Feynman amplitudes are constructed just from the free part, as it happens in standard electrodynamics in the non-linear gauge. Therefore, standard electrodynamics in the non-linear gauge $A_\mu A^\mu-b^2=0$ and the bumblebee model with solutions restricted to the condition $B_\mu B^\mu-b^2=0$ are described by identical Feynman rules and propagators, making both theories equivalent. Note that the interpretations of particles and constrains are different; the masslessness of photons is explained by gauge symmetry in standard electrodynamics, while in the bumblebee model is derived from the nature of Goldstone bosons arising from a spontaneous symmetry breaking of Lorentz invariance. The constraints $A_\mu A^\mu-b^2=0$ and $B_\mu B^\mu-b^2=0$ have the same structure; however, they have different origins, it is a gauge fixing condition in standard electrodynamics, while in the bumblebee model is just a restriction over its solutions.

\acknowledgments

We thank R. Potting, J. P. Noordmans and L. F. Urrutia for many valuable discussions, comments, and suggestions. This work has been partially supported by the project DGAPA-UNAM, Project No. IN-104815 and the CONACyT Project No. 237503. C. A. E. is supported by CONACyT Posdoctoral Grant No. 234745.

\end{document}